\documentclass{ws-procs9x6}

\newcommand{\beq}{\begin{equation}}
\newcommand{\eeq}{\end{equation}}
\newcommand{\beqa}{\begin{eqnarray}}
\newcommand{\eeqa}{\end{eqnarray}}
\newcommand{\no}{\nonumber}
\newcommand{\Lagr}{\mathcal{L}}

\begin{document}

\vspace{-1.0cm}
\hfill{\tiny HISKP-TH-06/35, FZJ-IKP-TH-2006-28}

\title{SOME RECENT DEVELOPMENTS IN\\
CHIRAL PERTURBATION THEORY\footnote{Work supported in part by
DFG (TR-16), by BMBF (06BN411) and by EU I3HP (RII3-CT-2004-506078).}}

\author{Ulf-G. Mei{\ss}ner$^\ddagger$}

\address{HISKP, Universit\"at Bonn, D-53115 Bonn, Germany\\
and\\
IKP, Forschungszentrum J\"ulich, D-52425 J\"ulich, Germany\\
$^\ddagger$E-mail: meissner@itkp.uni-bonn.de}

\begin{abstract}
In this talk, I address some recent developments in chiral perturbation theory
at unphysical and physical quark masses.
\end{abstract}


\bodymatter

\section{Introduction I: Remarks on chiral extrapolations}
\label{sec:intro1}

The first part of this talk concerns the application of chiral
perturbation theory (CHPT) at {\bf unphysical} quark masses.
More precisely, lattice QCD (LQCD) allows one in principle to calculate
hadronic matrix elements ab initio using capability computing on
a discretized space-time. To connect to the real world, various
extrapolations are necessary: LQCD operates at a finite volume $V$, 
at a finite lattice spacing $a$ and at large (unphysical) quark masses $m_q$.
All these effects can be treated in suitably tailored effective 
field theories (EFTs) (for a recent review, see~\cite{Sharpe:2006pu}).
Here, I consider the quark mass expansion of certain baryon
observables. Various nucleon (baryon) observables  
have already been computed on the lattice, like e.g.
masses of ground and excited states, magnetic moments, nucleon
electromagnetic radii, the nucleon axial-vector coupling, and so on.
CHPT in principle provides extrapolation functions for all these
observables, parameterized in terms of a number of low-energy constants
(LECs).  These LECs relate many observables, they are {\em not} 
dependent on the process one considers.
Given the present situation with only a few lattice results
at reasonably small quark masses available, it is mandatory 
to  incorporate as much phenomenological input as is available for the LECs
from studies of pion-nucleon scattering, pion production and so on.
Also, since many LECs appear in various observables, a true check of
our understanding of the chiral symmetry breaking of QCD requires
global fits at sufficiently small quark masses. It is absolutely
necessary for such extrapolations to make sense that one is in 
a regime where higher order terms stay sufficiently small. Consequently,
results must be independent of the regularization scheme,
these can differ by higher order terms (e.g. comparing results
based on heavy baryon CHPT to ones obtained employing e.g. infrared 
regularization). For this interplay of CHPT and LQCD to make sense,
the lattice ``data'' should be in the {\em true} chiral regime.
I will  illustrate these issues for two very different examples:
the axial-vector coupling $g_A$ and the Roper mass $m_R$.

\section{Application I: The nucleon axial-vector coupling}

\begin{figure}[tb]
\centerline{
\epsfig{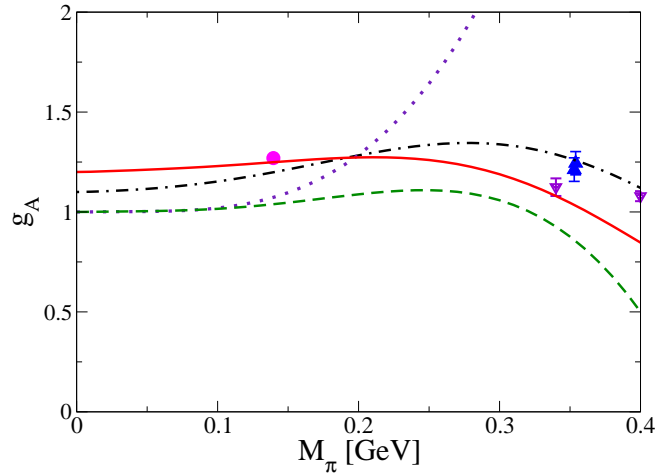}
}
\vspace{0.1cm}
\begin{center}
\caption{The axial-vector coupling $g_A$ as a function of the pion mass.
The solid (red), the dot-dashed (black) and the dashed (green) line
correspond to  various input values for the LECs 
(see \protect\cite{Bernard:2006te} for details).
The (magenta) circle denotes the physical value of $g_A$ at the 
physical pion mass, the triangles are the lowest mass data from  
Ref.~\protect\cite{Edwards:2005ym} and the
inverted triangles are recent results from QCDSF \protect\cite{Meinulf}.
\label{fig:gA}}
\vspace{-0.7cm}
\end{center}
\end{figure}

The nucleon axial-vector coupling $g_A$ is a fundamental quantity
in hadron physics as it appears prominently in the Goldberger-Treiman
relation. Lattice results for $g_A$ obtained by various collaborations
for pion masses between 300 and 1000~MeV show a very flat quark (pion)
mass dependence. On the the other hand, it is known since long that the one--loop
representation of $g_A$ is not converging well and is dominated 
by the $M_\pi^3$ term with increasing pion mass\cite{Kambor:1998pi}, thus
$g_A (M_\pi)$ rises steeply as the pion mass increases.
The large coefficient of this term is, however, understood in terms
of the large values of the dimension two LECs $c_3$ and $c_4$ combined
with some large numerical prefactors.  In Ref.~\cite{Bernard:2006te} 
we have therefore worked out the two--loop representation of $g_A$,
\begin{eqnarray}
\label{gAstruc}
g_A &=& g_0 \,\, \biggl\{ 1 + \left( \frac{\alpha_2}{(4\pi F)^2} \ln
\frac{M_\pi}{\lambda} + \beta_2 \right) \, M_\pi^2 + \alpha_3 \, M_\pi^3
\nonumber\\
+ && \!\!\!\!\! \left(\frac{\alpha_4}{(4\pi F)^4} \ln^2\frac{M_\pi}{\lambda}
+  \frac{\gamma_4}{(4\pi F)^2} \ln\frac{M_\pi}{\lambda} + \beta_4
\right) \, M_\pi^4 + \alpha_5 \, M_\pi^5 \biggr\} + {\cal O}(M_\pi^6)~,
\nonumber \\
&=&  g_0 \,\, \biggl\{ 1 + \Delta^{(2)} + \Delta^{(3)} + \Delta^{(4)} +
    \Delta^{(5)} \biggr\}   + {\cal O}(M_\pi^6)~,
\end{eqnarray}
with $g_0$ the chiral limit value of $g_A$, $\lambda$ is the scale of
dimensional regularization, and the coefficients $\alpha_{2,3}, \beta_2$
encode the one-loop result. Further, $F$ denotes the chiral limit value of
the pion decay constant, $F \simeq 86\,$MeV, and $\Delta^{(n)}$ collects
the corrections $\sim M_\pi^n$.
At two-loop order, one has corrections of fourth
and fifth order in the pion mass, given in terms of the coefficients
$\alpha_{4,5}, \beta_4, \gamma_4$ (note that there is also a $M_\pi \ln
M_\pi^5$ terms whose contribution we have absorbed in the uncertainty of
$\alpha_5$). The LEC $\alpha_4$ can be analyzed using the renormalization
group (as stressed long ago by Weinberg \cite{Weinberg:1978kz}) and is
entirely given in terms of the dimension three coefficients of the 
one--loop generating functional. We have performed this calculation 
based on two existing versions of the dimension three pion-nucleon chiral 
Lagrangian without and with equation of motion terms (see \cite{Ecker:1995rk} 
and \cite{Fettes:1998ud}, respectively) and obtained
\begin{equation}
\label{eq:inteomtot}
\alpha_4 = -\frac{16}{3} - \frac{11}{3} g_0^2 + 16 g_0^4~. 
\end{equation}
In \cite{Bernard:2006te} the dominant contributions to the LECs 
$\alpha_{5}, \beta_4, \gamma_4$ from $1/m_N$ corrections to dimension
two and three insertions  (with $m_N$ the nucleon mass)
as well as from the pion mass expansion of $F_\pi$
were also worked out, for details see that paper. 
Setting the remaining contributions to zero, 
we find (in the notation of  Eq.~(\ref{gAstruc}))
$g_A = g_0 ( 1 - 0.15 + 0.26 - 0.06 -0.001)$ and $g_0 = 1.21$,
using $g_A = 1.267$ and central values for the LECs $c_3, c_4, d_{16}$, see
e.g. \cite{Meissner:2005ba}. This shows that for the physical pion mass,
the higher order corrections are small and one thus has a convergent
representation. Varying the LECs $\alpha_{5}, \beta_4, \gamma_4$
within bounds given by naturalness, one finds that the pion mass dependence
of $g_A$ stays flat for $M_\pi \lesssim 350\,$MeV, see Fig.~\ref{fig:gA}.
\noindent From this figure one also sees that there is just a little overlap
between the lattice results and the chiral representation, where it can be
applied with a tolerable uncertainty. We remark that another solution to
this problem was offered in Ref.~\cite{Hemmert:2003cb} where an effective field theory
with explicit delta degrees of freedom at leading one--loop order
could lead to a flat pion mass dependence of $g_A$,
requiring, however, a  fine tuning of certain low--energy constants.
For a recent update, see~ \cite{Procura:2005ev}. It should also be noted
that most of the lattice results analyzed in these papers are far outside the
range of applicability of that particular EFT evaluated only to leading
one-loop order -- that such a representation works at such large pion masses
is an interesting observation but certainly does not support
claims of a controlled and precise determination of $g_A$ from LQCD.

\section{Application II: Chiral corrections to the Roper mass}

Understanding the (ir)regularities of the light quark baryon spectrum poses an 
important challenge for lattice QCD. In particular, the first even-parity
excited state of the nucleon, the Roper $N^* (1440)$ (from here on called the
Roper) is very intriguing---it is lighter than the first odd-parity nucleon
excitation, the $S_{11} (1535)$, and also has a significant branching ratio 
into two pions. Recent lattice studies have not offered a clear picture about 
the nucleon resonance spectrum.  In particular, in  Ref.~\cite{Mathur:2003zf} 
an indication of  a rapid cross over of the first positive and negative
excited nucleon states close to the chiral limit was reported -- so far not 
seen in other simulations at higher quark masses. Note also that so far
very simple chiral extrapolation  functions have been employed in most 
approaches, e.g., a linear  extrapolation in the quark masses, 
thus  $\sim M_\pi^2$, was applied in \cite{Burch:2006cc}. It is therefore 
important to provide the lattice practitioners with improved chiral 
extrapolation functions. A complete one--loop representation for the pion mass
dependence of the Roper mass was recently given in \cite{Borasoy:2006fk}.
Since the Roper is the first even-parity excited state of the nucleon, the
construction of the chiral SU(2) effective Lagrangian follows standard
procedures,  see e.g.~\cite{Fettes:2000gb}. The effective Lagrangian 
relevant for our  calculation is (working in the isospin limit $m_u = m_d$ 
and neglecting electromagnetism)
\beqa  \label{eq:Lagrangian}
\Lagr &=& \Lagr_0 + \Lagr_{R} + \Lagr_{NR}~, \no \\
\Lagr_0 &=& i \bar{N} \gamma_\mu D^\mu N - m_N \bar{N} N 
     +  i \bar{R} \gamma_\mu D^\mu R - m_R \bar{R} R ~,\no\\
\Lagr_{R}^{(1)} &=& \frac{1}{2} g_R \bar{R} \gamma_\mu \gamma_5 u^\mu R~, ~~
\Lagr_{NR}^{(1)} = \frac{g_{NR}}{2} \bar{R} \gamma_\mu \gamma_5 u^\mu N
               + {\rm h.c.}~, \no\\
\Lagr_{R}^{(2)} &=&  c_1^* \langle \chi_+ \rangle \bar{R}  R  
               - \frac{c_2^*}{8 m_R^2}  \bar{R} \left(\langle u_\mu u_\nu \rangle 
                    \{D^\mu, D^\nu \} + {\rm h.c.} \right)  R
               + \frac{c_3^*}{2}  \langle u_\mu u^\mu \rangle \bar{R}  R~ \ ,
\no
\eeqa
\beqa
\Lagr_{R}^{(4)} &=&  - \frac{e_1^*}{16} \langle \chi_+ \rangle^2 \bar{R}  R  \ ,
\eeqa
where $N, R$ are nucleon and Roper fields, respectively, and $m_N, m_R$
the corresponding baryon masses in the chiral limit. 
The pion fields are collected in $u_\mu = - \partial_\mu 
\mbox{\boldmath$ \pi$} / F_\pi  + {\cal O}(\mbox{\boldmath$ \pi$}^3)$.
$D_\mu$ is the chiral covariant derivative, for our purpose we can set
$D_\mu = \partial_\mu$, see e.g.~\cite{Fettes:2000gb} for definitions. 
Further, $\chi_+ $ is proportional to the pion mass and induces explicit chiral 
symmetry breaking, and $\langle ~ \rangle$ denotes the trace in 
flavor space. The dimension two and four LECs $c_i^*$ and $e_i^*$ correspond
to the $c_i$ and $e_i$  of the effective chiral pion--nucleon Lagrangian.
The pion-Roper coupling  is given  to leading chiral order by
$\Lagr_{R}^{(1)}$, with a coupling $g_R$.
This coupling  is bounded by the nucleon axial coupling, $|g_R| < |g_A|$,
in what follows we use $g_R = 1$. The leading interaction piece between 
nucleons and the Roper is given by $\Lagr_{NR}^{(1)}$.
The coupling $g_{NR}$ can be determined from the strong decays of the
Roper resonance, its actual value is $g_{NR} = 0.35$ using the Roper width
extracted from the speed plot (and not from a Breit-Wigner fit). 
Further pion-Roper couplings are encoded in $\Lagr_{R}^{(2)}$ and
$\Lagr_{R}^{(4)}$.  To analyze the real part of the Roper self-energy,
one has to calculate a) tree graphs with insertion $\sim c_1^*, e_1^*$,
self-energy diagrams with intermediate b) nucleon and c) Roper states and d)
tadpoles with vertices from $\Lagr_{R}^{(2)}$. In fact, the graphs of type b)
require a modification of the regularization scheme due to the appearance of
the two large mass scales $m_N$ and $m_R$. The solution to this problem --
assuming $m_N^2/m_R^2 \ll 1$ (in nature, this ratio is $\simeq 1/2.4$) -- 
is described in \cite{Borasoy:2006fk}. As discussed in that paper, the LECs
$c_i^*$ and $e_i^*$ can be bounded assuming naturalness and by direct
comparison with the corresponding pion-nucleon couplings: 
$|c_1^*| \lesssim 0.5\,$GeV$^{-1}$, $|c_{2,3}^*| \lesssim 1.0\,$GeV$^{-1}$
and $|e_1^*| \lesssim 0.5\,$GeV$^{-3}$. 
\begin{figure}[t]
\centering
\includegraphics[width=0.7\textwidth]{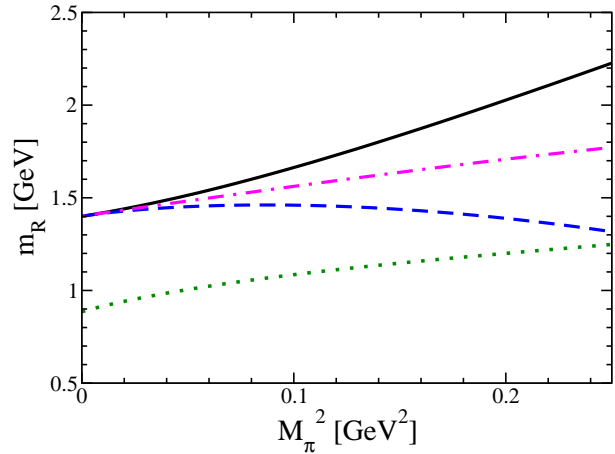} 
\caption{Quark mass dependence of the Roper mass for different parameter sets 
$c_1^* = -0.5, c_{2,3}^*, e_1^*$. The $c_i$ are in units of GeV$^{-1}$ and 
$e_1$ is given in GeV$^{-3}$.  
and couplings $g_R = 1.0, g_{NR} = 0.35$. The solid curve corresponds to
$c_2^* = 1.0, c_3^* = 1.0, e_1^* = 0.5$, the dashed one to 
$c_2^* = -1.0, c_3^* = -1.0, e_1^* = -0.5$ and the dot--dashed one to
$c_2^* = c_3^* = e_1^* = 0$. The dotted curve represents the quark mass 
dependence of the nucleon, see Ref.~\cite{Bernard:2003rp}. The values of the
corresponding LECs are: $c_1 = -0.9, c_2 = 3.2, c_3 = -3.45,e_1 = -1.0$.
}
\label{fig:mro}
\end{figure}
\noindent
In Fig.~\ref{fig:mro} an estimated range for the pion mass dependence of the Roper mass
is presented by taking the extreme values for $c_{2,3}^*$ and $e_1^*$, while keeping
$c_1^* = -0.5\, {\rm GeV}^{-1}, g_{NR} = 0.35, g_R = 1$ fixed. 
The masses of the baryons in the chiral limit
are taken to be $m_N = 0.885\,$GeV \cite{Bernard:2003rp}
and $m_R = 1.4\,$GeV, respectively.
The dash-dotted curve is obtained by setting the couplings $c_{2,3}^*,e_1^*$
all to zero, and exhibits up to an offset
a similar quark mass dependence as the nucleon result (dotted curve,
taken from Ref.~\cite{Bernard:2003rp}). 
It should be emphasized, however, that the one-loop formula cannot be trusted
for pion masses much beyond 350 MeV. No sharp decrease of the Roper mass
for small pion masses is observed for natural values of the couplings.
Note that the important $\Delta \pi $ and $N \pi\pi$ channels are
effectively included through the dimension two and four contact interactions,
still it would be worthwhile to extend these considerations including the
delta explicitely. Note further that the formalism developed in
\cite{Borasoy:2006fk} is in general suited to study systems with two heavy
mass scales in addition to a light mass scale. In this sense,
it can be applied to other resonances as well, such as the $S_{11}(1535)$. 
In this case, however, an SU(3) calculation is necessary due to the important 
$\eta N$ decay channel.

\section{Intoduction II: Hadronic atoms}

Let us come back to the real world of physical quark masses. In what
follows, I will discuss a spectacular effect of isospin violation
in pionic deuterium, which is one particular hadronic atom. More generally,
such atoms are made of certain hadrons  bound by the static Coulomb force.
There exist many species, e.g. $\pi^+\pi^-$, $\pi^\pm K^\mp$, $\pi^-
  p$, $\pi^- d$, $K^- p$, or $K^- d$. In these systems, the Bohr radii are
much larger than any typical scale of strong interactions (QCD), so that their
effects can be treated as perturbations. These are the 
energy shift $\Delta E$ from the Coulomb value and the decay width $\Gamma$
(often combined in the complex valued energy shift). Since the average momenta 
in such systems are very small compared to any hadronic scale, hadronic atoms
give access to scattering at zero energy and thus the pertinent S-wave
scattering length(s).  As it is well known, these scattering lengths are very 
sensitive to the  chiral and isospin symmetry breaking in QCD.
In fact, hadronic atoms offer may be the most precise
method of determining these fundamental parameters since theory and experiment
can in some cases be driven to an accuracy of one or a few percent.
On the theoretical side, hadronic atoms can be analyzed  systematically 
and consistently in the framework of non-relativistic low-energy EFT 
including virtual photons, see e.g. \cite{Rusetsky:2000ft}. 

\section{Isospin violation in pionic deuterium}

\begin{figure}[t]
 \includegraphics[width=10.4cm]{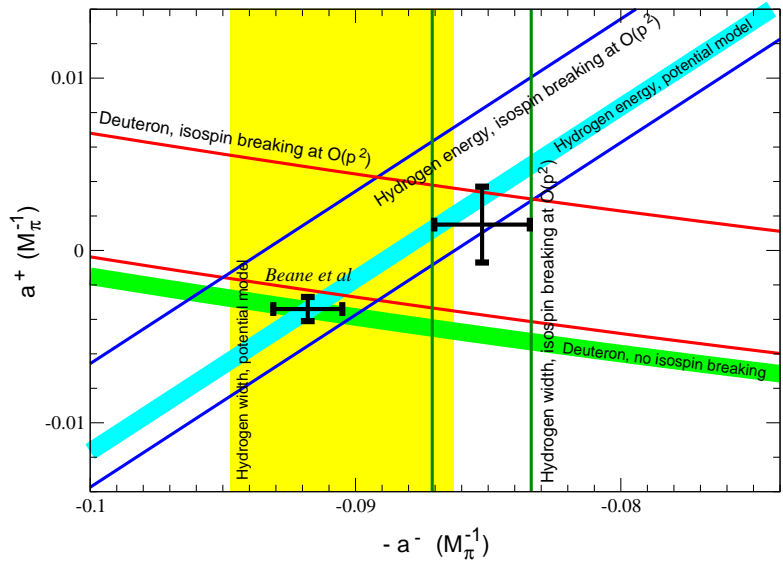}%
\vspace{0.2cm}
 \caption{Determination of the $\pi N$ $S$-wave scattering lengths
$a^+$ and $a^-$ from the combined analysis of the experimental 
data on the pionic hydrogen
energy shift and width, as well as
the pionic deuterium energy shift (details in the text).
The cross denoted as {\it Beane et al} is taken from 
Ref.~\cite{Beane:2002wk}. The second cross corresponds to
the scattering lengths given in Eq.~(\ref{eq:scattl}).
\label{fig:piatom}}
 \end{figure}

Combined measurements of the energy shift and decay width of pionic hydrogen and the
energy shift of pionic deuterium offer an excellent test of isospin symmetry
and its breaking because these three quantities are expressed in terms of two
scattering lengths, $\Delta E(\pi^- p) \sim a^+ + a^-$, 
$\Gamma (\pi^- p) \sim a^-$, and   ${\rm Re}\,\Delta E(\pi^- d) \sim a^+ +
\ldots$, where the ellipses denotes three-body effects such as multiple
scattering within the deuteron. Here, $a^+$ and $a^-$ are the isoscalar and
the isovector S-wave $\pi N$ scattering lengths, respectively.
The Bern group has championed the EFT
treatment of pionic hydrogen, the calculations including strong and
electromagnetic isospin violation can be found in \cite{Gasser:2002am} and
\cite{Zemp} for the ground state energy and the width, respectively. Using
this formalism to analyze the data from PSI (as reviewed in \cite{Gotta:2005cg})
and combining these with the
EFT treatment of pionic deuterium in the isospin limit \cite{Beane:2002wk}
(compare the bands denoted hydrogen energy, isospin breaking and hydrogen width,
isospin breaking and deuteron, no isospin breaking in Fig.~\ref{fig:piatom})
one faces a problem - these bands do not intersect. Since the analysis of pionic
deuterium in Ref.~ \cite{Beane:2002wk} was done in the isospin limit,
the question naturally arises whether this is the source of the trouble? In
principle, isospin violation (IV) leads to corrections in the bound-state
as well as in the pertinent scattering amplitude. From experience with 
pionium, pionic hydrogen and $\pi K$ atoms 
(see e.g. Ref.\cite{Schweizer:2004qe}), such 
bound-state corrections are expected to be small. On the other hand,
already in 1977 Weinberg pointed out that IV effects can be unnaturally
large  if the isospin-conserving (IC) contribution is chirally
suppressed \cite{Weinberg:1977hb}. In particular, such an effect
is very pronounced in neutral pion scattering off nucleons
(for an update, see \cite{Fettes:1998wf}), but it is very hard to observe.
On the other hand, the leading order contribution to $\pi d$ scattering is
chirally suppressed, ${\rm Re}\, a_{\pi d} \sim (a_{\pi^- p} + a_{\pi^- n})
= {\cal O}(p^2)$, despite the fact that $a_{\pi^- p}$ and $a_{\pi^- n}$ are
individually of ${\cal O}(p)$.  Here, $p$ denotes collectively 
the small parameters of CHPT. While this is well-known, nobody has ever 
systematically investigated IV in $\pi^- d$.  The leading order IV in pionic
deuterium was only analyzed recently  in Ref.~\cite{Meissner:2005ne}. 
To be specific, consider the  threshold pion-deuteron scattering amplitude:
\beq
{\rm Re}~a_{\pi d}^{\rm thr} = {\rm Re}~a_{\pi d}^{(0)} + \Delta a_{\pi d}\, ,
\eeq
where the IV piece $\sim \Delta a_{\pi d}$ appears at the same order as
the  leading IC piece $\sim {\rm Re}~a_{\pi d}^{(0)}$.
Also, to this order one has  no dependence on the deuteron structure.
The explicit calculation leads to
\beq
\label{eq:LO}
\Delta a_{\pi d}^{\rm LO} = (4\pi(1+\mu/2))^{-1}(\delta {\mathcal T}_p
+\delta {\mathcal T}_n)\, .
\eeq
Here, $\mu = M_{\pi^+}/m_p$ and ${\mathcal T}_{p,n}$ are the leading
isospin breaking corrections to the $\pi^- p$ and $\pi^- n$ threshold
scattering amplitudes. These are given by
\beqa
\delta{\mathcal T}_p&=&\frac{4(M_{\pi^+}^2-M_{\pi^0}^2)}{F_\pi^2}\,c_1
-\frac{e^2}{2}\,(4f_1+f_2)+ {\cal O}(p^3)\, ,
\nonumber\\
\delta{\mathcal T}_n&=&\frac{4(M_{\pi^+}^2-M_{\pi^0}^2)}{F_\pi^2}\,c_1
-\frac{e^2}{2}\,(4f_1-f_2)+ {\cal O}(p^3)\, ,
\eeqa
where $F_\pi=92.4~{\rm MeV}$ is the pion decay constant,
$g_A=1.27$ denotes the axial-vector charge of the nucleon and
$c_1$ is a strong and $f_1,f_2$ are  electromagnetic
${\cal O}(p^2)$  LECs, respectively.
Note also that at lowest order in CHPT $c_1$ is directly related to 
the value of the pion-nucleon $\sigma$-term and $f_2$ to the proton-neutron
mass difference. In the numerical calculations we take
$c_1=-0.9^{+0.5}_{-0.2}~{\rm GeV}^{-1}$~\cite{Meissner:2005ba},
$f_2=-(0.97\pm 0.38)~{\rm GeV}^{-1}$ \cite{Meissner:1997ii,Gasser:2002am}. 
Note that the errors on the LEC $c_1$ are most conservative.
The largest uncertainty in the
results is introduced by the constant $f_1$, whose value at present
is unknown and for which  the dimensional estimate
$|f_1|\leq 1.4~{\rm GeV}^{-1}$ has been used. Note also, that the hydrogen
energy band, which is shown in Fig.~\ref{fig:piatom}
corresponds to the new value of $c_1$ given above.
From this, the leading order IV contribution to pionic deuterium follows as:
\beqa
\label{eq:LOpid}
\Delta a_{\pi d}^{\rm LO} &=& (4\pi(1+\mu/2))^{-1}(\delta {\mathcal T}_p
+\delta {\mathcal T}_n) \nonumber\\
&=& \frac{1}{4\pi (1 + \mu/2)} 
\left\{ \frac{8\Delta M_\pi^2}{F_\pi^2} \, c_1 - 4e^2 \, f_1 
\right\} + {\cal O}(p^3) \, ,
\eeqa
with $\Delta M_\pi^2 = M_{\pi^+}^2 - M_{\pi^0}^2$ the squared 
charged-to-neutral pion mass difference.
Substituting numerical values for the various low-ener\-gy constants, which
were specified above,
one obtains that the correction at ${\cal O}(p^2)$ is extremely large
\begin{equation}
\Delta a_{\pi d}^{\rm LO} = -(0.0110^{+0.0081}_{-0.0058}) \, M_\pi^{-1}~,
\end{equation}
that is
$\Delta a_{\pi d}^{\rm LO}/ {\rm Re}\,a_{\pi d}^{\rm exp}=0.42$ 
(central values), using the experimental value ${\rm Re}\,a_{\pi d}^{\rm exp}
= -(0.0261\pm 0.0005)M_\pi^{-1}$ \cite{Hauser:1998yd}.
Moreover, one can immediately see that
the correction moves the de\-u\-te\-ron band in 
Fig.~\ref{fig:piatom} in the right direction: the isospin-breaking 
corrections amount for the bulk of the discrepancy between the experimental
data on  pionic hydrogen and deuterium. 
Including the corrections $\Delta a_{\pi d}^{\rm LO}$, all bands now 
have a common 
intersection area in the $a^+,a^-$-plane, see Fig.~\ref{fig:piatom}.
The resulting values for the $\pi N$ scattering lengths are:
\begin{eqnarray}\label{eq:scattl}
a^+ &=& (0.0015\pm 0.0022) \, M_\pi^{-1}\, , \nonumber\\[2mm] 
a^- &=& (0.0852\pm 0.0018) \, M_\pi^{-1}\, .  
\end{eqnarray}
Further, using the hydrogen energy shift to estimate the LEC $f_1$, we obtain
\begin{equation}
f_1= -2.1^{+3.2}_{-2.2}~{\rm GeV}^{-1}\, ,
\end{equation}
which is consistent with the dimensional analysis and a recent evaluation
based on a quark model~\cite{Lyubovitskij:2001zn}.
Note that the error displayed here does not include the uncertainty coming
from the higher orders in CHPT and should thus be considered preliminary. 

As we see, the presence of the ${\cal O}(p^2)$ LECs in the expressions for the
isospin-breaking corrections leads to a sizeable increase of the uncertainty
in the output. In order to gain  precision, 
in the fit one might also use those particular linear
combination(s) of the experimental observables that 
do not contain $f_1$ and $c_1$. However, it should be pointed out that
such a fit imposes much more severe constraints on 
the data than the uncorrelated fit considered above. 
In fact, applying isospin-breaking 
corrections only at ${\cal O}(p^2)$, we find that the data are still over-constrained 
in the combined fit. From this we finally conclude that
 to carry out such a combined analysis with the required precision,
one would have e.g. first to evaluate the  isospin-breaking corrections 
with a better accuracy.

Up to now, we have restricted ourselves to the leading-order isospin-breaking
correction in CHPT. Calculations at ${\cal O}(p^3)$ exist only for the hydrogen
energy shift and yield 
$\delta_\epsilon=(-7.2\pm 2.9)\cdot 10^{-2}$~\cite{Gasser:2002am} (using
$c_1=(-0.93\pm 0.07)~{\rm GeV}^{-1}$).
The corrections to $O(p^2)$ result are sizable
(the energy band in Fig.~\ref{fig:piatom} will be shifted further upwards), 
but the uncertainty, which is almost completely determined by the 
${\cal O}(p^2)$ LECs, remains practically the same.
On the other hand, consistent studies at ${\cal O}(p^3)$ 
imply the treatment of the scattering process in the
three-body system in the effective field theory with virtual photons. 
To the best of our knowledge, such investigations have not been yet carried 
out, although certain three-body contributions at ${\cal O}(p^3)$  were
calculated in the past~\cite{Rockmore:1995vi}. It is natural to expect 
that generally three-body terms at ${\cal O}(p^3)$ should not depend on the 
additional LECs from the two-nucleon 
sector and hence the extraction of the $\pi N$ scattering lengths at a high
precision is still possible. Of course, these arguments
can not be a substitute for a rigorous proof in the framework
of EFT, which in the light of the above discussion, is urgently called for.
It was also shown in  Ref.~\cite{Meissner:2005ne} that the ${\cal O}(p^4)$
correction which emerges from the double-scattering term in the 
multiple-scattering series is very small, $\Delta a_{\pi d}^{\rm double~scat.}
= 0.003\, {\rm Re}\,a_{\pi d}^{\rm exp}\,$, using the  scattering lengths 
from Eq.~(\ref{eq:scattl}) as input. Of course, such a partial result can 
only be considered indicative. Evidently, the systematic analysis 
of all ${\cal O}(p^3)$ (and eventually ${\cal O}(p^4)$) corrections should 
be carried out.

\section{Summary and outlook}

In the first part of this talk, I have considered aspects related
to baryon CHPT for light  quark masses above their physical
values. It should be stressed that
baryon CHPT is a mature field in the up and down quark sector and
provides unambiguous extrapolation functions for LQCD -- certainly
more work is needed for the three-flavor case. To carry out the
required chiral extrapolations, one should keep in mind that different
observables are linked by general operator structures and the appearing 
low-energy constants (LECs) are universal, which means that they
are independent of the process considered.
Given the present status of LQCD, it appears mandatory to perform 
global fits to observables and constrain the appearing LECs 
by input from phenomenology, whenever available. I have 
discussed two specific examples of the interplay
between CHPT and LQCD. A chiral extrapolation for $g_A$  exists now at 
two--loop accuracy. For pion masses $\lesssim 350$~MeV, the theoretical uncertainty
related to it is reasonably small.  I have also provided a 
chiral extrapolation function  for the Roper mass - certainly much more 
work is needed for such excited states from both CHPT and LQCD.
Evidently, we need more lattice ``data'' at low quark masses to really
perform precision studies.

In the second part of this talk, I returned to the real world (physical
quark masses) and considered hadronic atoms. These can be systematically analyzed in 
non-relativistic effective field theory including virtual photons. We have found
a very large isospin-violating effect in pionic deuterium at leading order.
That there is such an effect is not so surprising because the leading 
isospin-conserving contribution is chirally suppressed. What is surprising,
however, is the actual size of the effect and that it was only found recently.
Combining this with the information obtained from the analysis of the 
energy shift and width in pionic hydrogen, one is led to a consistent 
extraction of the S-wave $\pi N$ scattering lengths and can furthermore determine 
the electromagnetic LEC $f_1$. Clearly, higher order calculations are  
necessary to reduce the theoretical  uncertainty.
In this context it is also important to stress that it was recently shown that
there are only {\em tiny} dispersive corrections to ${\rm Re}~a_{\pi d}$, 
see \cite{Lensky:2006wd} and Hanhart's talk at this conference.

\section*{Acknowledgements}
I thank the organizers for a superbe job and all my collaborators for sharing
their insight into the topics discussed here.

\end{document}